\documentclass[a4paper,11pt,dvips]{article}
\usepackage{graphicx}
\usepackage{amsmath}
\usepackage{amssymb}
\usepackage{latexsym}
\usepackage{makeidx}
\usepackage{color}

\DeclareMathOperator{\sech}{sech}
\newcommand{\bra}{\begin{array}}
\newcommand{\era}{\end{array}}
\newcommand{\beq}{\begin{equation}}
\newcommand{\eeq}{\end{equation}}
\newcommand{\bqr}{\begin{eqnarray}}
\newcommand{\eqr}{\end{eqnarray}}

\def\BC{\bb C}
\def\_\BC{\bbi C}

\def\( {\left(}
   \def\) {\right)}
\def\[ {\left[}
\def\] {\right]}
\def\no2 {{\textstyle{n\over 2}}}

\def\dag {{\dagger}}

\newcommand{\om}{\omega}

\newcommand{\lam}{\lambda}
\newcommand{\si}{\sigma}

\newcommand{\va}{\varepsilon}
\newcommand{\be}{\beta}

\newcommand{\del}{\delta}

\newcommand{\da}{\dagger}

\newcommand{\lb}{\label}

\begin{document}
\begin{titlepage}
\setcounter{page}{1}
\renewcommand{\thefootnote}{\fnsymbol{footnote}}

\begin{flushright}
ucd-tpg:1103.02\\
\end{flushright}

\vspace{5mm}
\begin{center}

{\Large \bf {Graphene Nanoribbon in Sharply Localized Magnetic Fields}}

\vspace{5mm}
{\bf A.D. Alhaidari}$^{a}$, {\bf H. Bahlouli}$^{a,b}$, {\bf A. El Mouhafid}$^{a,c}$ and
{\bf A. Jellal}$^{a,c,d}$\footnote{\sf ajellal@ictp.it - a.jellal@ucd.ac.ma}

\vspace{5mm}

{$^{a}$\em Saudi Center for Theoretical Physics, Dhahran, Saudi Arabia}

{$^{b}$\em Physics Department, King Fahd University of Petroleum $\&$ Minerals,  \\
Dhahran 31261, Saudi Arabia}

{$^{c}$\em Theoretical Physics Group,  
Faculty of Sciences, Choua\"ib Doukkali University},\\
{\em 24000 El Jadida,
Morocco}

{$^d$\em Physics Department, College of Sciences, King Faisal University,\\
Alahssa 31982, Saudi Arabia}

\vspace{3cm}

\begin{abstract}
We study the effect of a sharply localized magnetic field on the electron {transport}
in a strip (ribbon) of graphene sheet,
which allows to give results for the transmission and reflection probability through magnetic barriers.
The magnetic field is taken as a single and double delta type localized functions, which are
 treated later as the zero width
limit of gaussian {fields.} For both field configurations, we
{evaluate}
analytically and numerically
their transmission and reflection coefficients. The possibility of spacial confinement due to the inhomogeneous field
configuration is also investigated.

\end{abstract}
\end{center}

\vspace{3cm}

\noindent PACS numbers: 72.80.Vp, 73.21.-b, 71.10.Pm, 03.65.Pm

\noindent Keywords: graphene, scattering, pseudo-scalar potential, bound states.

\end{titlepage}

\section{Introduction}

Graphene, a single sheet of pure carbon atoms, has been the subject of intense focus and research for the last few years due to its unique and amazing properties that could make it more suitable for future electronic devices and possibly replace silicon-based semiconductors in electronics \cite{1,2}. Graphene nanoribbons, in particular, are of interest because they exhibit a conduction bandgap that arises due to size confinement and edge effects \cite{3}-\cite{11}. From the electronic applications point of view, graphene-based materials with gap are very desirable. This lead to substantial theoretical and experimental efforts to generate and control the energy gap in graphene based devices. One approach is based on quantum confinement as in quantum dots and nanoribbons, where it was shown that the energy gap value increases with decreasing nanoribbon width \cite{12}.

Theoretical studies have suggested that graphene nanoribbons could have interesting magneto-electronic properties, with predicted very large magnetoresistance \cite{4,13}. Experimental observation of a significant enhancement in the conductance of a graphene nanoribbon field-effect transistor by a perpendicular magnetic field has been reported recently \cite{14}. They have observed an extraordinarily large tunable magnetoresistance in graphene nanoribbon--FET devices and an increase in conductance by a factor of over 10,000 was demonstrated at $1.6 K$. This magnetoresistance can be readily tuned by the gate voltage and source-drain bias, with enhancement reaching maxima near the edges of the conduction bandgap. These experimental findings clearly demonstrate that the graphene nanoribbons exhibit interesting magneto-transport properties, with the possibility of opening up exciting opportunities in magnetic sensing and a new generation of magneto-electronic devices.

On the other hand, more efforts have been {made} to understand the scattering behavior of fermions
in graphene. In particular we cite \cite{peeters} where
the transmission through magnetic barriers in graphene-based nanostructures {were studied}. Several particular
cases were considered: a magnetic step, single and double barriers, and $\del$-function barriers. A separate class of
magnetic-barrier structures are those with inhomogeneous magnetic-field profiles, such that the average magnetic
field vanishes, which can be realized by nanostructured ferromagnetic stripes placed on top of the
graphene layer. Quantum bound states that are localized near or in the barrier are predicted for a magnetic step
and some structures with finite-width barriers but none for $\del$-function barriers. When a bound state is localized
close to the barrier edge, it has a nonzero velocity parallel to this edge. The transmission depends strongly on
the direction of the incident electron or hole wave vector and gives the possibility to construct a direction dependent wave vector filter. In general, the resonant structure of the transmission is significantly more
pronounced for (Dirac) electrons with linear spectrum than for the usual electrons with a parabolic spectrum.

Motivated by different results on the scattering behavior, we study
fermions in graphene nanoribbon subject to a sharply localized magnetic field.
{We show} that the solution space of the present system is the union {of two} solution subspaces
governed by two different kinetic balance equations. These two solution subspaces are not identical and their union constitutes the whole
space of solution {of the} problem. To {know} which energy domain is selected by each
kinetic balance relation, we perform a unitary transformation on the fundamental Dirac
equation. Such a transformation does not affect the energy spectrum or the physics of {the problem.}
After
matching the eigenspinors  at the boundary conditions, we evaluate the reflection and transmission
amplitudes in terms of the energy and magnetic field. A detailed study of their coefficients is
presented as well.

Subsequently, we consider a magnetic field {of a bell shape} normal to the graphene {strip,
which is stretched along the $x$-axis}, i.e. $\vec B(x)= \hat z B_0\sech^2(\mu x)$
where $B_0$ is the strength of the field and $\mu^{-1}$ {represents} its range. We show that the solution
space of the corresponding Hamiltonian splits into three regions depending on the sign of $kB_0$ and
the energy range, with $k$ {being the} wave vector along the $y$-axis. Using the corresponding current density, we evaluate the
reflection and transmission coefficients, which reduce to those obtained in section 3 for {very large $\mu$}. Finally, we treat
the case of a double step pseudoscalar potential and also double Gaussian type magnetic field.

{The} manuscript is organized as follows. We present our theoretical model by setting
{up} the general form of potential
in section 2.  As a first {illustration,} we consider $\delta$-type magnetic field in section 3.
After {obtaining the solutions of} the energy spectrum, we study the scattering problem by evaluating explicitly
the reflection and transmission coefficients and {investigate} their physical properties. Because {the} $\delta$-type magnetic field
is a {limiting case of a sharply peaked field}, we study Dirac fermions in a Gaussian magnetic field, in section 4.
These {results motivated us} to deal with an integrating case {of}  a double $\delta$-type magnetic field
in section 5 and the corresponding double Gaussian type magnetic field is considered in
{the} appendix. We conclude our work in the final section.


\section{Theoretical model}

In graphene, the low energy linear electronic band dispersion near the Dirac points gave rise to charge carriers (electrons or holes) that propagate as if they were massless fermions with speeds of the order of $10^6 m/s$ rather than the speed of light $3\times10^8 m/s$. Hence, charge carriers in this system should be described by the massless Dirac equation rather than the usual Schrodinger equation. However, we should stress that such a continuum approximation remains valid up to wave vectors of the order of the inverse lattice spacing. Under this approximation we can describe our charge carriers by a gauge invariant Dirac Hamiltonian in $2+1$ space-time dimensions with minimal coupling to the three-vector potential $A_{\mu } =(A_{0} ,\vec{A})$ $=(V,A_{x} ,A_{y} )$ as follows
\begin{equation} \label{GrindEQ1}
H=m\gamma _{0} +V+\gamma _{0} \vec{\gamma }\cdot (i\vec{\nabla }+\vec{A})
\end{equation}
where $\gamma _{0} =\sigma _{3} $, $\vec{\gamma }=(\gamma _{x} ,\gamma _{y} )=i(\sigma _{1} ,\sigma _{2} )$ and $\left\{\sigma _{i} \right\}_{i=1}^{3} $ are the 2$\times$2 Pauli matrices. Adopting the conventional relativistic units $\hbar =c=e=1$, we obtain \cite{15}
\begin{equation} \label{GrindEQ2}
H=\left(\begin{array}{cc} {m+V} & {-\frac{\partial }{\partial x} +i\frac{\partial }{\partial y} +iA_{x} +A_{y} } \\
{\frac{\partial }{\partial x} +i\frac{\partial }{\partial y} -iA_{x} +A_{y} } & {-m+V}
\end{array}\right).
\end{equation}

As we claimed before, for a system made of graphene the charge carriers are massless ${m} = 0$ and  moving with the Fermi velocity $v_{\sf F}$
(about $10^6 m/s$). Moreover, we specialize to a physical configuration corresponding to a single graphene sheet in a magnetic field with vector potential
$\vec{A}=\hat{y} W(x)$. Thus, the associated stationary Dirac equation becomes
\begin{equation} \label{GrindEQ3}
\left(\begin{array}{cc} {-\varepsilon } & {-\frac{\partial }{\partial x} +{\rm i}\frac{\partial }{\partial y} +W} \\ {\frac{\partial }{\partial x} +{\rm i}\frac{\partial }{\partial y} +W} & {-\varepsilon } \end{array}\right)\left(\begin{array}{c} {\mathop{\psi ^{+} }\limits^{} } \\  {\mathop{\psi ^{-} }\limits_{} } \end{array}\right)=0.
\end{equation}
We can separate variables and write
\beq
\psi ^{\pm } (x,y)=\varphi ^{\pm } (x)\chi (y)
\eeq
 where $\chi (y)=e^{-{\rm i}{\kern 1pt} ky} $ and \textit{k} is a real parameter that stands for the wave number of the excitations along the \textit{y}-axis. The resulting reduced equation is given by
\begin{equation} \label{GrindEQ4}
\left(\begin{array}{cc} {-\varepsilon } & {-\frac{d}{dx} +W+k} \\ {\frac{d}{dx} +W+k} & {-\varepsilon } \end{array}\right)\left(\begin{array}{c} {\varphi ^{+} } \\  {\varphi ^{-} } \end{array}\right)=0.
\end{equation}
In the present representation $W$ plays the role of a pseudoscalar potential coupling. The above equation is equivalent to the following set of equations
\begin{eqnarray}
&&\varphi ^{\mp } =\frac{1}{\varepsilon } \left(\pm \frac{d}{dx} +W+k\right)\varphi ^{\pm }\label{GrindEQ5}\\
&&\left[-\frac{d^{2} }{dx^{2} } +W^{2} \mp \frac{dW}{dx} +2kW+k^{2} -\varepsilon ^{2} \right]\varphi ^{\pm } =0.\label{GrindEQ6}
\end{eqnarray}
The structure of the Schr\"odinger-like equation (\ref{GrindEQ6}) has the features associated with supersymmetric quantum mechanics
(SSQM) \cite{16}. The two potential partners in SSQM are ${\rm {\mathcal V}}^{2} \pm {\rm {\mathcal V}}'$, where ${\rm {\mathcal V}}$
is the superpotential
 and in our case we have ${\rm {\mathcal V}}=W+k$. The relativistic extension of SSQM exhibited in some models that are governed by the Dirac theory is due to the fact that the associated Dirac Hamiltonian is an element of the 4-dimensional superalgebra $U(1/1)$, which is a $Z_{2} $ graded extension of $SO(2,1)$ Lie algebra \cite{17}. It is interesting to note that the corresponding massive (1+1) Dirac equation
\begin{equation} \label{GrindEQ7}
\left(\begin{array}{cc} {m-\varepsilon } & {-\frac{d}{dx} +W} \\ {\frac{d}{dx} +W} & {-m-\varepsilon } \end{array}\right)\left(\begin{array}{c} {\varphi ^{+} } \\  {\varphi ^{-} } \end{array}\right)=0
\end{equation}
results in the following Schr\"odinger-like equation
\begin{equation} \label{GrindEQ8}
\left[-\frac{d^{2} }{dx^{2} } +W^{2} \mp \frac{dW}{dx} +m^{2} -\varepsilon ^{2} \right]\varphi ^{\pm } =0
\end{equation}
which has the same structure associated with SSQM as in (\ref{GrindEQ6}) with
the mapping $W\to W+k$. Therefore, the solution of the two problems are similar with $k$ playing the role of the mass $m$. We have encountered this phenomenon before, where space reduction (due to compactification or symmetry) transforms  the 2D massless problem into an effective 1D massive problem \cite{18}. Note that the complete solution space of the Dirac equation (\ref{GrindEQ4}) is the union of the two solution subspaces of equations (\ref{GrindEQ5}) and (\ref{GrindEQ6})
with the top and bottom signs, respectively. Below, we identify the energy regions associated with each subspace.

\section{Dirac Fermions in a $\delta$-type magnetic field}

Here we consider the zero range and infinite strength limit where
the magnetic field takes the form $\vec{B}(x)=\hat{z} B_{0} \delta ({x\mathord{\left/ {\vphantom {x a}} \right. \kern-\nulldelimiterspace} a} )$, with \textit{a} is a length scale parameter. The main advantage of this configuration  is that it is amenable to analytical treatment
and interesting results. Thus,
the single-step pseudoscalar potential is given by
\beq
W(x)=a B_{0} \Theta ({x\mathord{\left/ {\vphantom {x a}} \right. \kern-\nulldelimiterspace} a} )
\eeq
 where $\Theta (x)$ being the Heaviside step function, such as
\begin{equation} \label{GrindEQ9}
\Theta (x)=\left\{\begin{array}{cc} {1}, & \qquad {x>0} \\ {0}, & \qquad {x<0} \end{array}\right.
\end{equation}
Figure 1 shows the schematics of a laboratory set-up where a coil of a rectangular cross section with very narrow width will result in the necessary current sheet that should produce a magnetic field approximating that of our problem. The graphene ribbon is the strip that penetrates the coil along the \textit{x}-axis in the figure:

\begin{center}
  \includegraphics[width=4in]{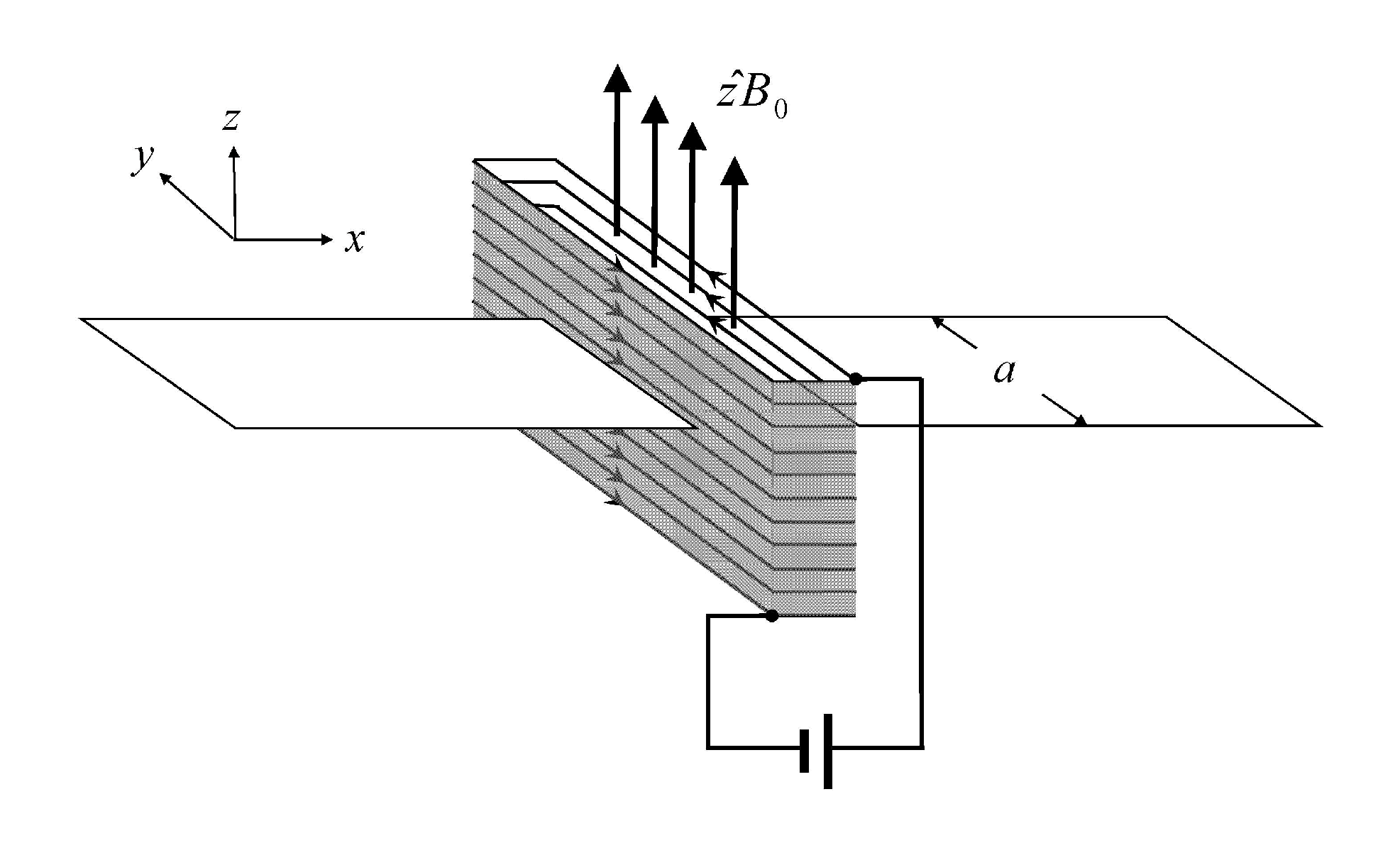}
\end{center}
{\sf{Fig. 1:} \sf{Schematic of a set-up where a coil of a rectangular cross section with very narrow width produces a magnetic field approximating that of our problem.}}\\

To study the scattering behavior in the present problem, we first determine the plane wave solutions associated with both regions, $\pm x>0$. The two components of the wavefunction satisfying equations (\ref{GrindEQ5}) and  (\ref{GrindEQ6}) with the top sign are
given by
\bqr
\varphi ^{-} =\frac{1}{\varepsilon } \left({+} \frac{d}{dx} +\omega _{\pm } \right)\varphi ^{+}           \lb{5p}\\
\left[-\frac{d^{2} }{dx^{2} } +\omega _{\pm }^{2} -\varepsilon ^{2} \right]\varphi ^{+} =0         \lb{6p}
\eqr
with $\omega _{+} =aB_{0} +k$ and $\omega _{-} =k$. It is easy to show that the solution of these equations
reads
\begin{equation} \label{GrindEQ10}
\varphi (x)=A\left(\begin{array}{c} {1} \\ {\alpha _{\pm } } \end{array}\right)e^{\kappa _{\pm } x} +B\left(\begin{array}{c} {1} \\ {\beta _{\pm } } \end{array}\right)e^{-\kappa _{\pm } x}
\end{equation}
  where the parameters read as
  \beq
  \kappa _{\pm } =\sqrt{\omega _{\pm }^{2} -\varepsilon ^{2} }, \qquad
  \alpha _{\pm } ={\left(\omega _{\pm } {+}\kappa _{\pm } \right)\mathord{\left/ {\vphantom {\left(\omega _{\pm } -\kappa _{\pm } \right) \varepsilon }} \right. \kern-\nulldelimiterspace} \varepsilon }, \qquad
  \beta _{\pm } ={\left(\omega _{\pm } {-}\kappa _{\pm } \right)\mathord{\left/ {\vphantom {\left(\omega _{\pm } +\kappa _{\pm } \right) \varepsilon }} \right. \kern-\nulldelimiterspace} \varepsilon }.
  \eeq
  On the other hand, considering the lower sign in equations (\ref{GrindEQ5}) and  (\ref{GrindEQ6}) for $\pm x>0$ we obtain
\bqr
 \varphi ^{+} =\frac{1}{\varepsilon } \left({-}\frac{d}{dx} +\omega _{\pm } \right)\varphi ^{-}           \lb{GrindEQ5pp}\\
\left[-\frac{d^{2} }{dx^{2} } +\omega _{\pm }^{2} -\varepsilon ^{2} \right]\varphi ^{-} =0        \lb{GrindEQ6pp}
\eqr
which can be solved to give
\begin{equation} \label{GrindEQ101}
\varphi (x)=A\left(\begin{array}{c} {\beta _{\pm } } \\ {1} \end{array}\right)e^{\kappa _{\pm } x} +B\left(\begin{array}{c} {\alpha _{\pm } } \\ {1} \end{array}\right)e^{-\kappa _{\pm } x}.
\end{equation}
In summary, the oscillatory solutions (for both cases and for $\pm x>0$) are obtained for all $\left|\varepsilon \right|>\left|\omega _{\pm }\right| $ while evanescent wave solutions will result if $\left|\varepsilon \right|<\left|\omega _{\pm }\right| $. Thus the quantity $\left|\omega _{\pm }\right|$ plays the role of an effective mass with different asymptotic values at $\pm x>0$. That is, our massless problem is equivalent to a space dependent massive problem whose effective mass depends on {the} strength of the localized magnetic field and the value of component of the wave number along the $y$-direction.

We need to reassert that the total solution space of the problem is the union of the two solution subspaces (\ref{GrindEQ5}) and (\ref{GrindEQ6}). These two solution subspaces are{, in general,} not identical and their union constitutes the whole space of solution of our problem. To resolve the issue of which energy domain is selected by which kinetic balance relation (\ref{5p}) or (\ref{GrindEQ5pp}), we perform a unitary transformation on the fundamental Dirac equation (\ref{GrindEQ4}). Such a transformation does not affect {the physics} of the problem. We choose a rotation by ${\pi \mathord{\left/ {\vphantom {\pi  4}} \right. \kern-\nulldelimiterspace} 4} $ about the $y$-axis, $U=e^{i{\textstyle\frac{\pi }{4}} \sigma _{2} } $. Thus, the transformed Hamiltonian and wavefunction
read
\begin{eqnarray} \label{GrindEQ12}
\tilde{H}&=&UHU^{\dag } =-i\sigma _{2} \frac{d}{dx} +(W+k)\sigma _{3}\\
  \tilde{\varphi }&=&U\varphi \nonumber
\end{eqnarray}
 then (\ref{GrindEQ4}) becomes
\begin{equation} \label{GrindEQ13}
\left(\tilde{H}-\varepsilon {\mathbb I}\right)\tilde{\varphi }=\left[-i\sigma _{2} \frac{d}{dx} +(W+k)\sigma _{3} -\varepsilon {\mathbb I}\right]\tilde{\varphi }
\end{equation}
 leading to the following new kinetic balance relation
\begin{equation} \label{GrindEQ14}
\tilde{\varphi }^{\mp } =\frac{\mp 1}{\varepsilon \pm (W+k)} \frac{d}{dx} \tilde{\varphi }^{\pm }
\end{equation}
which is not valid for
\beq
\varepsilon =\mp \left(aB_0+k\right)=\mp \om_+.
\eeq
Thus, {for $k> -aB_0$}, this relation with the top/bottom sign corresponds to the positive/negative energy subspace\footnote{
{However, if $k< -aB_0$} then the top/bottom sign corresponds to negative/positive energy {subspace.}}. 
Now, since equation (\ref{GrindEQ13}) results from a unitary transformations of (\ref{5p}) and (\ref{GrindEQ5pp}), then the same applies to those equations as well. That is, equations (\ref{5p}) and  (\ref{GrindEQ5pp}) are valid for positive and negative energy, respectively. Figure 2 shows the potential configuration of the problem. Dark/light grey areas correspond to negative/positive energy continuum where the solution is oscillatory of the form $e^{\pm i\kappa x} $. The white areas correspond to bound states where the solutions are exponentials, $e^{\pm \kappa x} $. In the subsequent developments, we will only be considering scattering solutions with a normalized incident beam from left. Thus, the relativistic energy must be $\varepsilon >\left|k\right|$ and the solution space is built only from the positive energy solutions (\ref{GrindEQ10}).

\begin{center}
  \includegraphics[width=2.8in]{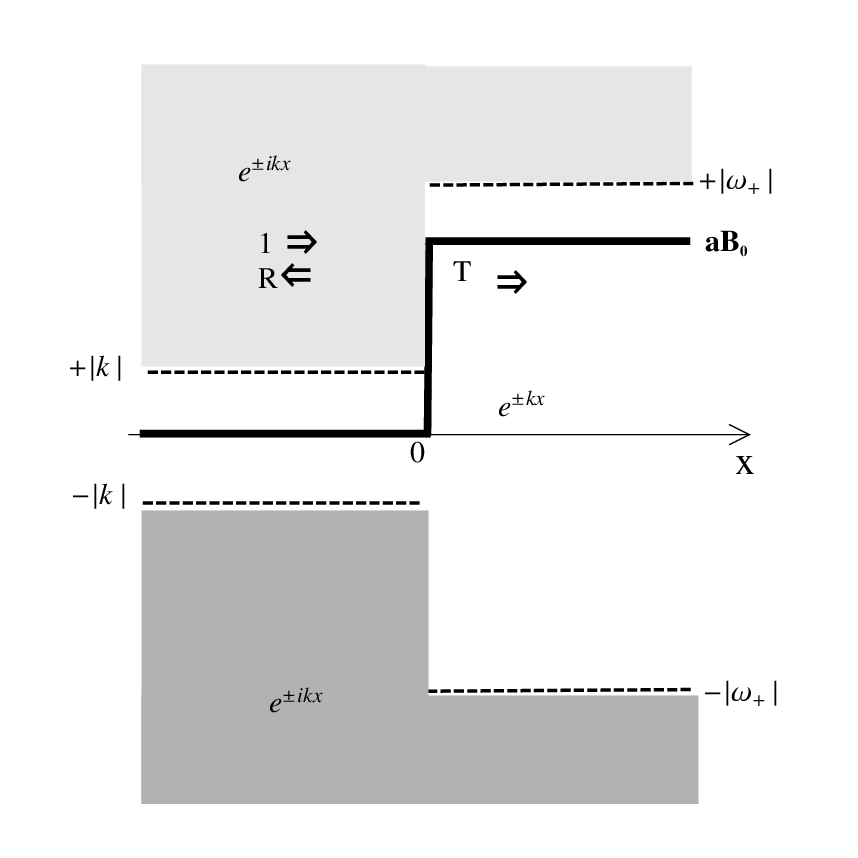}
\end{center}
{\sf{Fig. 2:}  \sf{The potential configuration of the problem. Dark/light grey areas correspond to negative/positive energy continuum. The white areas correspond to bound states where the solutions are exponentials.}}\\

The total solution space splits into two regions depending on the energy range as follows. For {$|k|<\varepsilon <|\omega _{+}|$}, scattering with total reflection will take place. It should be noted that there is no Klein tunneling in this problem due to the absence of vector interaction. With a normalized incident flux from left, the solution of equation (\ref{GrindEQ4}) is obtained from (\ref{GrindEQ10}) as
\bqr
\varphi (x)=\left\{\begin{array}{cc} {\left(_{\alpha }^{\, 1} \right)e^{i\kappa x} +R\left(_{\alpha ^{*} }^{\, \, 1} \right)e^{-i\kappa x} }, & \qquad {x<0} \\ {A\left(_{\beta _{+} }^{\, \, 1} \right)e^{-\kappa _{+} x} }, & \qquad {x>0} \end{array}\right.
\eqr
where {we have set
\beq
\kappa =\left|\kappa _{-} \right|=\sqrt{\varepsilon ^{2} -k^{2} }, \qquad \alpha ={(k-i\kappa )/\varepsilon }.
\eeq}
Continuity of the wavefunction at \textit{x} = 0 gives the reflection amplitude
\begin{equation}\lb{RRT}
R=-\frac{\beta _{+} -\alpha }{\beta _{+} -\alpha ^{*} } =-\frac{aB_{0} +\kappa _{+} +i\kappa}{aB_{0} +\kappa _{+} -i\kappa}
\end{equation}
and the normalization parameter $A=1+R$. It is clear that we have complete reflection in this case since $\left|R\right|=1$. It is also worth mentioning that the probability density in the positive $x$-direction is given by
\beq
\rho (x) = |A|^2 e^{-2 |\kappa _{+ }| x}(1+|\beta _{+ }|^2)\lb{ro}
\eeq
which shows that the penetration depth, $\lambda$, of the Fermion into the barrier region defined by
\beq
\lambda = \frac{1}{2|\kappa _{+ }|^2}=\frac{1}{2\sqrt{(aB_0+k)^2-\epsilon^2}}
\eeq
decreases very rapidly with increase in the magnetic field strength. This result indicates the possibility to confine the fermions in any region between two
$\del$-like magnetic fields generated by a symmetric double step pseudoscalar potential, this situation will be treated in section 5.

Now, let us move on to analyze the scattering situation when {$\varepsilon >|\omega _{+}|$} so that partial reflection will take place. Again, with a normalized incident wave from left, the solution of  (\ref{GrindEQ4}) is obtained from (\ref{GrindEQ10}) as
\bqr
\varphi (x)=\left\{\begin{array}{cc} {\left(_{\alpha }^{\, 1} \right)e^{i\kappa x} +R\left(_{\alpha ^{*} }^{\, \, 1} \right)e^{-i\kappa x} }, & \qquad {x<0} \\ {T\left(_{\beta }^{\, 1} \right)e^{i\gamma x} }, & \qquad {x>0} \end{array}\right.
\eqr
{with the quantities
 \beq
 \gamma =\left|\kappa _{+} \right|=\sqrt{\varepsilon ^{2} -\omega _{+}^{2} }, \qquad
 \beta ={\left(\omega _{+} -i\gamma \right)\mathord{\left/ {\vphantom {\left(\omega _{+} -i\gamma \right) \varepsilon }} \right. \kern-\nulldelimiterspace} \varepsilon }.
 \eeq}
 Again matching the spinor wavefunctions
 at \textit{x} = 0 gives the reflection and transmission amplitudes
\beq
R=-\frac{\alpha -\beta }{\alpha ^{*} -\beta }, \qquad T=\frac{{2i\kappa \mathord{\left/ {\vphantom {2i\kappa  \varepsilon }} \right. \kern-\nulldelimiterspace} \varepsilon } }{\alpha ^{*} -\beta } \lb{rtt}.
\eeq
One can verify the flux conservation relation
\beq
\left|R\right|^{2} +\left|{\gamma \mathord{\left/ {\vphantom {\gamma  \kappa }} \right. \kern-\nulldelimiterspace} \kappa } \right|\left|T\right|^{2} =1.
\eeq
Figure 3 is a plot of the reflection and transmission coefficients as a function of the
energy for several values of the magnetic field strength $B_{0} $.\\

\begin{center}
  \includegraphics[width=3.2in]{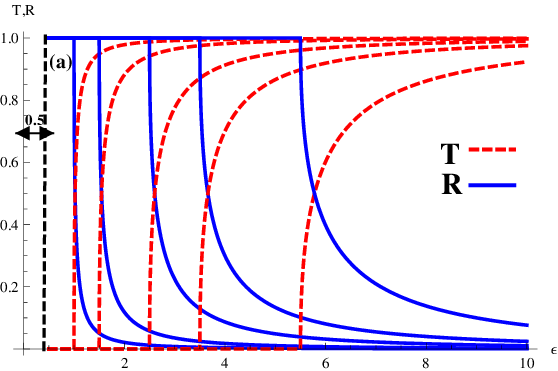}\ \ \ \  \includegraphics[width=3.2in]{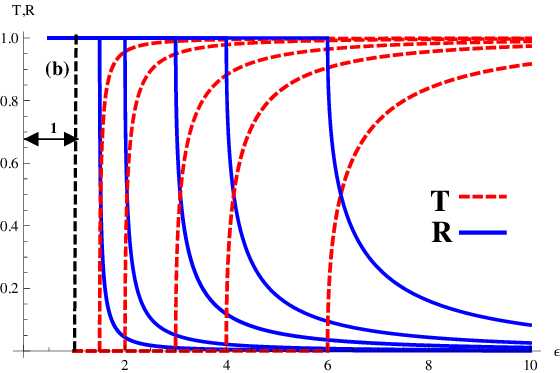}
 \end{center}
  {\sf{Fig. 3:  \sf{Reflection and transmission amplitudes as a function of energy for $B_{0}=0.5, 1.0, 2.0, 3.0, 5.0$ in units of $a^{-2}$ (left to right curves) {where (a) for $k=0.5$ and (b) for $k=1$}. We defined $\mathbf{\mathrm{T}}(\varepsilon)=T(\varepsilon)\sqrt{\gamma/\kappa}$.}}}\\

\noindent It is also worthwhile to study the transmission coefficient as a function of the magnetic field strength $B_{0} $. For this purpose we can write explicitly $|T|$, in this regime, as follows
\beq
|T| = 2\sqrt{\frac{\va^2 -k^2} {(aB_0)^2 + \left( \sqrt{\va^2 -k^2} +\sqrt{\va^2 - (aB_0+k)^2}\right)^2}}
\eeq
which shows clearly that the transmission is unity at $B_0=0$ and reduces to zero for strong field.
Figure 4 shows the behavior of the transmission and reflection coefficients as a function of the applied magnetic field. We would like to note that the transmission vanishes for energies below the critical energy value of {$\varepsilon^{\ast}=|\om_+|=B_{0}+k$}, as shown in figure 3. A similar remark can be made regarding figure 4.

\begin{center}
  \includegraphics[width=3.5in]{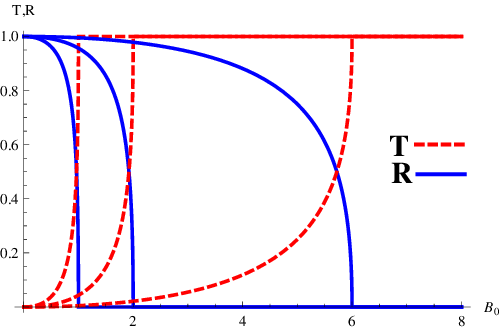}
 \end{center}
{\sf{Fig. 4:  {Reflection and transmission amplitudes as a function of strength of the field for $\varepsilon=3, 4, 8$ in units of $a^{-2}$ (left to right curves) and $k=2$.}}}

\section{Dirac fermions in a Gaussian magnetic field}

We consider a magnetic field normal to the graphene sheet of a bell shape. Specifically,
we take $\vec{B}(x)=\hat{z}B_{0}\sech^{2}(\mu x)$ (Fig. 5)~\cite{kuru,milpas}, where $B_{0} $ is the strength of the field and $\mu ^{-1} $ stands for its range. The corresponding vector potential (Fig. 5) is
\beq
\vec{A}(x)=\hat{y}W(x)= \hat{y} \left(B_{0} /\mu \right)\ \tanh (\mu x)
\eeq
 which  turns equation (\ref{GrindEQ6}) into
\begin{equation}
 \left[-\frac{d^{2} }{dx^{2} } \mp B_{0} \left(1-\tanh^{2} \mu x\right) +\left(\frac{B_{0} }{\mu } \tanh \mu x+k\right)^{2}-\varepsilon ^{2} \right]\varphi ^{\pm } =0.   \lb{A1}
\end{equation}
Note that as $x\to \pm \infty $, this equation reduces to a simple form
\beq
\left(-\frac{d^{2} }{dx^{2} } +\omega _{\pm }^{2} -\varepsilon ^{2} \right)\varphi ^{\pm } =0
\eeq
 where $\omega _{\pm } =k\pm B_{0} /\mu $. This equation is similar to that of a massive 1D problem with asymptotic masses $m=\left|\omega _{\pm } \right|$ at $x\to \pm \infty $. The solution space of  (\ref{A1}) splits into three regions depending on the sign of $kB_{0} $ and the energy range. If $kB_{0} >0$, then $\left|\omega _{-} \right|<\left|\omega _{+} \right|$
 and these energy regions can be defined as follows
\begin{itemize}
     \item  $\left|\varepsilon \right|<\left|\omega _{-} \right|$: Bound states with the boundary condition $\mathop{\lim }\limits_{x\to \pm \infty } \varphi (x)=0$.

     \item  $\left|\omega _{-} \right|<\left|\varepsilon \right|<\left|\omega _{+} \right|$: Scattering with total reflection to the negative \textit{x}-axis (note that there is no Klein tunneling in this problem due to the absence of vector interaction). The boundary conditions in this case are: $\mathop{\lim }\limits_{x\to +\infty } \varphi (x)=0$, $\mathop{\lim }\limits_{x\to -\infty } \varphi (x)=e^{{\rm i}{\kern 1pt} \kappa _{-} x} \left(_{\rho _{-} }^{{\kern 1pt} {\kern 1pt} {\kern 1pt} 1} \right)+{\rm {\mathcal R}}e^{-{\rm i}{\kern 1pt} \kappa _{-} x} \left(_{\rho _{-}^{*} }^{{\kern 1pt} {\kern 1pt} 1} \right)$, where $\kappa _{\pm } =\sqrt{\varepsilon ^{2} -\omega _{\pm }^{2} } $, $\rho _{\pm } =\frac{1}{\varepsilon } \left(\omega _{\pm } -{\rm i}{\kern 1pt} \kappa _{\pm } \right)$ and the reflection amplitude is such that $\left|{\rm {\mathcal R}}\right|=1$.

     \item  $\left|\varepsilon \right|>\left|\omega _{+} \right|$: Scattering solutions occur with the boundary conditions: $\mathop{\lim }\limits_{x\to +\infty } \varphi (x)=$ ${\rm {\mathcal T}}e^{{\rm i}{\kern 1pt} \kappa _{+} x} \left(_{\rho _{+} }^{{\kern 1pt} {\kern 1pt} {\kern 1pt} 1} \right)$, $\mathop{\lim }\limits_{x\to -\infty } \varphi (x)=e^{{\rm i}{\kern 1pt} \kappa _{-} x} \left(_{\rho _{-} }^{{\kern 1pt} {\kern 1pt} {\kern 1pt} 1} \right)+{\rm {\mathcal R}}e^{-{\rm i}{\kern 1pt} \kappa _{-} x} \left(_{\rho _{-}^{*} }^{{\kern 1pt} {\kern 1pt} 1} \right)$, where is
         ${\rm {\mathcal T}}$ the transmission amplitude such that $\left|{\rm {\mathcal R}}\right|^{2} +\left|\rho _{+} /\rho _{-} \right|\left|{\rm {\mathcal T}}\right|^{2} =1$. However, if $kB_{0} <0$, then $\left|\omega _{-} \right|>\left|\omega _{+} \right|$ and the above subspaces should be interchanged with $\omega _{+} \leftrightarrow \omega _{-} $ and $x\to -x$.\\
\end{itemize}

\begin{center}
 \includegraphics[width=2in]{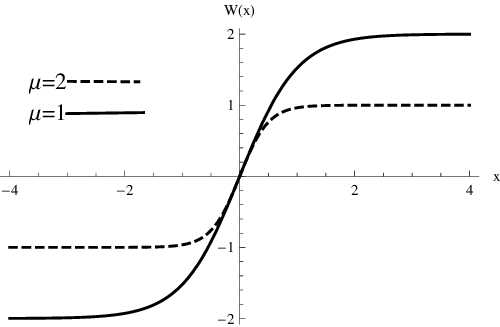}
  \includegraphics[width=2in]{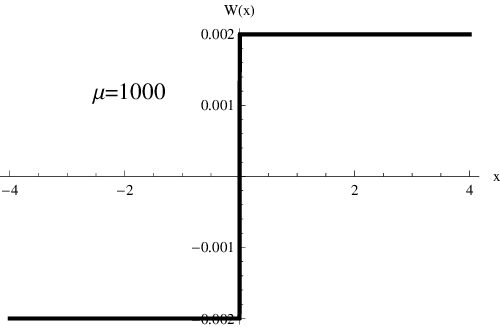}
  \includegraphics[width=2in]{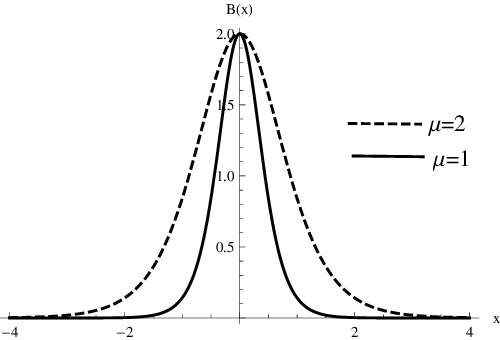}
  \end{center}
{\sf Fig. 5}:  {\sf{Hyperbolic tangent pseudoscalar potential profile along with the associated magnetic field for the strength $B_{0}=2$.}}\\

 To be more specific, let us  consider the scattering problem in the negative $x$ regions, $x<0$. Making the change of variable $z=-e^{2\mu x}$, equation (\ref{A1}) becomes
\begin{equation}
 \left[4 {\mu}^{2} z \frac{d}{dz}\left(z\frac{d}{dz}\right)\mp B_{0} \frac{4 z}{(1-z)^{2}} -\left(-\frac{B_{0} }{\mu } \frac{1+z}{1-z}+k\right)^{2}+\varepsilon ^{2} \right]\varphi ^{\pm} =0.\lb{AA}
\end{equation}
Now, we  solve (\ref{AA}) for $\varphi ^{+}$ and then deduce
$\varphi ^{-}$ using (\ref{GrindEQ5}). {In doing so,
we multiply  (\ref{AA}) by $\frac{1-z}{z}$ and make the change
$\varphi ^{\pm}= z^{h'}(1-z)^{\lam} F(z)$ to end up with}
{
\beq
\left[z(1-z) \frac{d^2}{dz^2} + \left( 2h' +1 - z(2h' +2\lam +1)\right)\frac{d}{dz}
- (h'+ \lam)^2 + \frac{1}{4\mu^2} \left(-\varepsilon^2 +\left(\frac{B_0}{\mu}+k\right)^2 \right)\right] F=0.
\eeq
This is nothing but the hypergeometric second order differential equation. Therefore,
the general solution for $\varphi^+$ can be expressed in terms of the hypergeometric functions~\cite{hf}}
\begin{eqnarray}
\varphi ^{+} &=&A z^{h'} (1-z)^{\lambda} {_{2}}{F}_{1}\left(h'-\nu+\lambda,h'+\nu+\lambda,1+2h;z\right)\nonumber\\
 &+&B z^{-h'} (1-z)^{\lambda}
{_{2}}{F}_{1}\left(-h'-\nu+\lambda,-h'+\nu+\lambda,1-2h';z\right)\lb{so}
\end{eqnarray}
where we have set
$h'=\frac{\sqrt{\omega_{-}^{2}-\varepsilon^{2}}}{2\mu}$,
$\nu=\frac{\sqrt{\omega_{+}^{2}-\varepsilon^{2}}}{2\mu}$ and
$\lambda=\frac{1}{2}-\frac{\sqrt{k^{2}+4B_{0}+4({B_{0}}/{\mu})^{2}}}{2\mu}$.
To study the scattering problem we need to investigate the
asymptotic behavior of (\ref{so}) when $x\rightarrow-\infty$,
which corresponds to $z\rightarrow 0$. Using
 the asymptotic behavior of the hypergeometric functions we obtain for $\varphi ^{+}$
\begin{equation}
\mathop{\lim }\limits_{x\to - \infty }\varphi ^{+}=A e^{ik_{-} x}+B e^{-ik_{-} x} \lb{s}
\end{equation}
with $k_{-}=\sqrt{\varepsilon^{2}-\omega_{-}^{2}}$. From
$(\ref{GrindEQ5})$ the other component, $\varphi ^{-}$, is  given
by
\begin{equation}
\varphi ^{-} =\frac{1}{\varepsilon } \left(\frac{d}{dx} +\frac{B_{0}}{\mu}+k\right)\varphi ^{+} .\lb{f}
\end{equation}
Substituting equation $(\ref{s})$ in $(\ref{f})$ gives the asymptotic behavior of $\varphi ^{-}$
\begin{equation}
\mathop{\lim }\limits_{x\to - \infty }\varphi ^{-}=A \eta_{+} e^{ik_{-} x}+B \eta_{-} e^{-ik_{-} x} \lb{}
\end{equation}
where $\eta_{\pm}=\frac{\omega_{+}\pm ik_{-}}{\varepsilon}$.

Next, we consider the solution of equation $(\ref{A1})$ for $x>0$. The analysis of the solution can be simplified by making the substitution $z=-e^{-2\mu x}$. Equation $(\ref{A1})$ becomes
\begin{equation}
 \left[4 {\mu^{2}} z \frac{d}{dz}\left(z\frac{d}{dz}\right)\mp B_{0} \frac{4 z}{(1-z)^{2}} -\left(\frac{B_{0} }{\mu } \frac{1+z}{1-z}+k\right)^{2}+\varepsilon ^{2} \right]\varphi ^{\pm} =0.\lb{Ab}
\end{equation}
The general solution to the above is
\begin{eqnarray}
\varphi ^{+} &=& C z^{h} (1-z)^{\lambda} {_{2}}{F}_{1}\left(h-\nu+\lambda,h+\nu+\lambda,1+2h;z\right)\nonumber\\
&+& D z^{-h} (1-z)^{\lambda} {_{2}}{F}_{1}
\left(-h-\nu+\lambda,-h+\nu+\lambda,1-2h;z\right)\lb{}
\end{eqnarray}
where $h=\frac{\sqrt{\omega_{+}^{2}-\varepsilon^{2}}}{2\mu}$,
$\nu=\frac{\sqrt{\omega_{-}^{2}-\varepsilon^{2}}}{2\mu}$ and
$\lambda=\frac{1}{2}-\frac{\sqrt{k^{2}+4B_{0}+4({B_{0}}/{\mu})^{2}}}{2\mu}$.

Also as $x\rightarrow-\infty$, $z\rightarrow 0$. Therefore in
order to have a plane wave travelling to the right as
$x\rightarrow \infty$, we require that $\left|\varepsilon
\right|>\left|\omega _{+} \right|$ and $C=0$ so that as
$x\rightarrow\infty$ we have
\begin{equation}
\varphi ^{+} =D z^{-h} (1-z)^{\lambda} {_{2}}{F}_{1}
\left(-h-\nu+\lambda,-h+\nu+\lambda,1-2h;z\right)\lb{}
\end{equation}
and the limit
\begin{equation}
\mathop{\lim }\limits_{x\to + \infty }\varphi ^{+}=D e^{ik_{+} x} \lb{ss}
\end{equation}
with $k_{+}=\sqrt{\varepsilon^{2}-\omega_{+}^{2}}$. The other
component is
\begin{equation}
\mathop{\lim }\limits_{x\to + \infty }\varphi ^{-}=D \beta
e^{ik_{+}x} \lb{sss}
\end{equation}
where $\beta=\frac{\omega_{+}+ik_{+}}{\varepsilon}$.

For one-dimensional scattering problems, the particles in the beam are in plane wave states with definite momentum.
Given the wave functions relevant to incident, reflected and transmitted beams, one may calculate the corresponding
current densities~\cite{DH}. The reflection and transmission coefficients are defined to be the ratio of the reflected current
density $j^{\sf re}$ to the incident current density $j^{\sf in}$ and transmitted current density $j^{\sf tr}$ to the incident current
density, respectively, {where the current density for our system is
$J= \varphi^{\da} \si_x \varphi$
with $\varphi= \left(\begin{array}{c} {\varphi ^{+} } \\  {\varphi ^{-} } \end{array}\right)$.
This is  leading to the coefficients}
\begin{equation}\label{RT}
R=\frac{\left|j^{\sf re}\right|}{\left|j^{\sf in}\right|}=
\frac{\left|\eta_{-}-\eta_{-}^{\ast}\right|}{\left|\eta_{+}-\eta_{+}^{\ast}\right|}\left|r\right|^{2},\\ \qquad
T=\frac{\left|j^{\sf tr}\right|}{\left|j^{\sf in}\right|}=
\frac{\left|\beta-\beta^{\ast}\right|}{\left|\eta_{+}-\eta_{+}^{\ast}\right|}\left|t\right|^{2}
\end{equation}
{where
\beq
r=\frac{\eta_+ -\be}{\be - \eta_-}, \qquad t=\frac{\eta_+ - \eta_-} {\be - \eta_-}
\eeq
which giving the conservation law $R+T=1$.}\\
\begin{center}
  \includegraphics[width=2.5in]{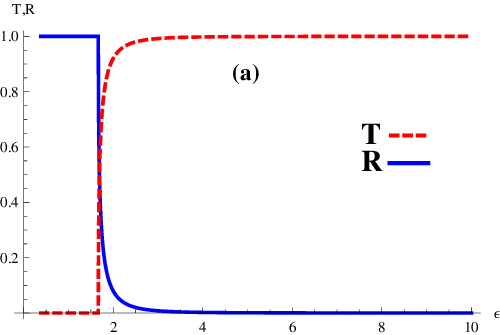}
   \includegraphics[width=2.5in]{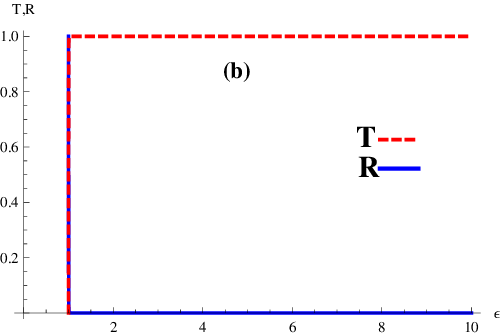}\\
{\sf Fig. 6}:  \sf{Transmission as a function of energy with $k=1$,  $B_{0} = 2$ and (a) $\mu=3$, (b) $\mu=1000$.}
\end{center}

\noindent We notice that for the Hyperbolic tangent magnetic field  (figure 5), figure 6 show clearly that for the value of $\mu=1000$ we obtain the same result for the transmission and reflection coefficients as those generated in section 3 for a single step pseudoscalar potential.

We can also characterize the transmission amplitude by giving the
corresponding phase shifts $\delta$. Indeed, in the case of $\left|\omega _{-} \right|<\left|\varepsilon \right|<\left|\omega _{+} \right|$
one obtains
\begin{equation}\label{psh}
t=\left|t\right|e^{i\delta}
\end{equation}
{where $|t|=\sqrt{T}$ given in \eqref{RT}}.
The calculations lead to the scattering phase shift in the form
\beq
\delta=\arctan\left(\frac{h}{h''}\right), \qquad h''=\frac{\sqrt{\varepsilon^{2}-\omega_{-}^{2}}}{2k}.
\eeq
A complementary information about the phase shift can be obtained by studying the trajectories in the complex
transmission amplitude plane where full reflections will appear whenever these trajectories pass through the origin.

\section{Dirac fermions in a double $\delta$-type magnetic field }

Motivated by the above analysis, let us investigate the scattering of Dirac fermions in a double $\delta$-type magnetic field. This can be described by
the double-step pseudoscalar potential
\beq \lb{double}
W(x)=B_1\left[1-\Theta(x+a)\right]+B_2\Theta(x-a)
\eeq
presented in figure 7  with $a$ defined to be positive real number.
This pseudoscalar double step potential generates a magnetic field $B = B_1 \delta(x+a) + B_2 \delta(x-a)$, $B_1$ and $B_2$ are the strengths of the  $\delta$-type magnetic fields
at $x = -a$ and $x = +a$, respectively.\\
\begin{center}
  \includegraphics[width=3in]{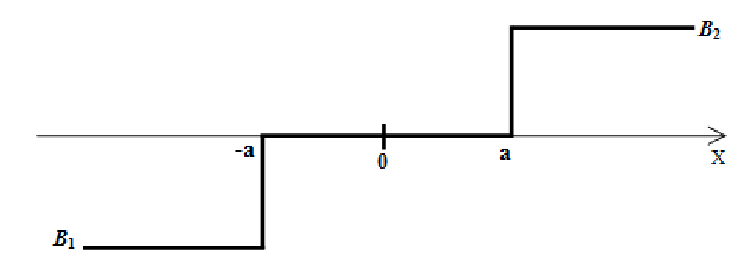}\\
{\sf Fig. 7}:  \sf{Double-step pseudoscalar potential.}
\end{center}

Using (\ref{double}), we show that the general solutions of the Dirac equation (\ref{GrindEQ6}) for the case of $\left|\varepsilon \right|>\left|\omega _{+} \right|$ can be written as
\begin{eqnarray}\label{41}
\begin{array}{l}
\ \ \ \ {\varphi (x<-a)=\left\{\begin{array}{cc} {A_-}e^{{\rm i}\, \kappa _{-} x} \left(_{\alpha _{-} }^{1} \right)+B_-e^{-{\rm i}\, \kappa _{-} x} \left(_{\alpha _{-}^{*} }^{1} \right), & \qquad \qquad \qquad \varepsilon <\omega _{+}  \\
A_+e^{{\rm i}\, \kappa _{-} x} \left(_{1}^{\alpha _{-}^{*} } \right)+{B_+}e^{-{\rm i}\, \kappa _{-} x} \left(_{\; 1}^{\, \alpha _{-} } \right), &
\qquad \qquad \qquad \ \   \varepsilon >-\omega _{+}
\end{array}\right. } \\
\ \ \ \ {\varphi (\left|x\right|<a)=\left\{\begin{array}{cc} {C_-}e^{{\rm i}\, k _{2} x} \left(_{\gamma _{+} }^{1} \right)+D_-e^{-{\rm i}\, k _{2} x} \left(_{\gamma _{+}^{*} }^{1} \right), & \qquad \qquad\qquad \ \   \varepsilon <\omega _{+}  \\
{C_+}e^{{\rm i}\, k _{2} x} \left(_{1}^{\gamma _{+}^* } \right)+{D_+}e^{-{\rm i}\, k _{2} x} \left(_{1}^{\gamma _{+} } \right), &
 \qquad\qquad\qquad \ \  \ \ \  \varepsilon >-\omega _{+}  \end{array}\right.}  \\
 \ \ \ \ \ \ {\varphi (x>a)=\left\{\begin{array}{cc} {E_-}e^{{\rm i}\, \kappa _{+} x} \left(_{\beta _{+} }^{1} \right)+F_-e^{-{\rm i}\, \kappa _{+} x} \left(_{\beta _{+}^{*} }^{1} \right), & \qquad \qquad\qquad \  \varepsilon <\omega _{+}  \\
 {E_+}e^{{\rm i}\, \kappa _{+} x} \left(_{1}^{\beta _{+}^* } \right)+{F_+}e^{-{\rm i}\, \kappa _{+} x} \left(_{1}^{\beta _{+} } \right), &
 \qquad\qquad\qquad \ \ \ \varepsilon >-\omega _{+}  \end{array}\right.}  \end{array}
 \end{eqnarray}
where {we define
\beq
 \gamma _{+}=(ik_{2}-k)/\varepsilon, \qquad
 \kappa_\pm=\sqrt{\varepsilon^2-\omega _{\pm}^2}, \qquad
 k_2=\sqrt{\varepsilon^2-k^2}, \qquad
 \omega _{+}=B_{2}+k, \qquad\omega _{-}=B_{1}+k.
 \eeq}
 These wavefunctions will be used to study the scattering and the bound states for the present case.

 The continuity of the wavefunction at $x=\pm a$ gives the reflection amplitudes $r_\mp=B_{\mp}/A_{\mp}$ and the transmission amplitudes $t_\mp=E_{\mp}/A_{\mp}$ for the negative (top sign) and positive (bottom sign) energy solutions. More precisely, we find
\begin{eqnarray}
 t_{\mp}&=&\frac{4e^{-ia(-2k_{2}+\kappa_-+\kappa_+)}\kappa_-k_{2}}{e^{4iak_{2}}(\kappa_- -k_{2}\mp iB_{1})(k_{2}-\kappa_+ \mp iB_{2})
 +(\kappa_- +k_{2}\mp iB_{1})(k_{2}+\kappa_+ \pm iB_{2})}\label{tt}\\
r_{\mp}&=&\frac{e^{-2ia\kappa_-}\left[e^{4iak_{2}}(\kappa_- +k_{2}\mp iB_{1})(k_{2}-\kappa_+ \mp iB_{2})+(\kappa_- -k_{2}
\pm iB_{1})(k_{2}+\kappa_+ \pm iB_{2})\right]}{e^{4iak_{2}}(\kappa_- -k_{2}\mp iB_{1})(k_{2}-\kappa_+ \mp iB_{2})
 +(\kappa_- +k_{2}\mp iB_{1})(k_{2}+\kappa_+ \pm iB_{2})}.\label{rr}
\end{eqnarray}
The result of the passage to the limits, $B_{1}\rightarrow 0$ ($k_{2}\rightarrow\kappa_-$) and $a\rightarrow0$, in  (\ref{tt}) and (\ref{rr}) coincides with the transmission and reflection amplitudes in (\ref{rtt}) of the Dirac particle scattered by a pseudoscalar single step potential.  If we take the limit $a \rightarrow 0$ in the above equations we get, by replacing $\omega_+$ and $\omega_-$ by $\omega_+ = B_2-k$ and $\omega_- = B_1-k$, respectively, and $B_0$ by $B_1-B_2$, the transmission and reflection amplitude for the Dirac particle scattered by a double-step pseudoscalar potential    $W(x)$ (\ref{double}). 
For the special case $B_2 = -B_1=0.002$ the transmission and reflection coefficients coincide with those obtained in the case of scattering of Dirac fermions by a single step pseudoscalar potential plotted in figure 6 (b).

Given the wavefunctions associated with the incident, reflected and transmitted beams, one may calculate
the corresponding current densities according to~\cite{DH}. Figure 8 illustrate the configuration of the solution
space of our problem. Dark/light grey areas correspond to negative/positive energy continuum where we have
evanescent or transmitted waves (EWs/TWs). The white areas correspond to the bound states (BS).
The reflection and transmission coefficients for the positive and negative energy solutions are
\begin{equation}\label{wx}
{R_{\mp}}=\left|r_{\mp}\right|^{2}, \qquad
{T_{\mp}}=\frac{\Re(\kappa_+)}{\Re(\kappa_-)}\left|t_{\mp}\right|^{2}.
\end{equation}
which obey the conservation relation
\beq
 \left|r_{\mp}\right|^{2} + \frac{\Re(\kappa_+)}{\Re(\kappa_-)}\left|t_{\mp}\right|^{2}=1.
\eeq
 Figures 9 and 10 show more clearly that, for the confinement by a double-step pseudoscalar potential, the EWs, TWs and BS always appear for the energy that $\left|\omega _{-} \right|<\left|\varepsilon \right|<\left|\omega _{+} \right|$ , $\left|\varepsilon \right|>\left|\omega _{+} \right|$ and $\left|\varepsilon \right|<\left|\omega _{-} \right|$, respectively. In figures 9 (a), (b) and 10 (b) the transmission resonances appear for $B_1$ and $B_2\neq0$. From figure 10 (a) one readily notice that, for the value of $B_2=0$, the double-step pseudoscalar potential reduces to a one step with $W(x)=B_1(1-\Theta(x))$. We can easily find the EWs, TWs and BS for this later representation by replacement the $\omega_+$ and $\omega_-$ by $\omega_+=k$ and $\omega_-=B_1-k$ respectively.
\begin{center}
  \includegraphics[width=2.6in]{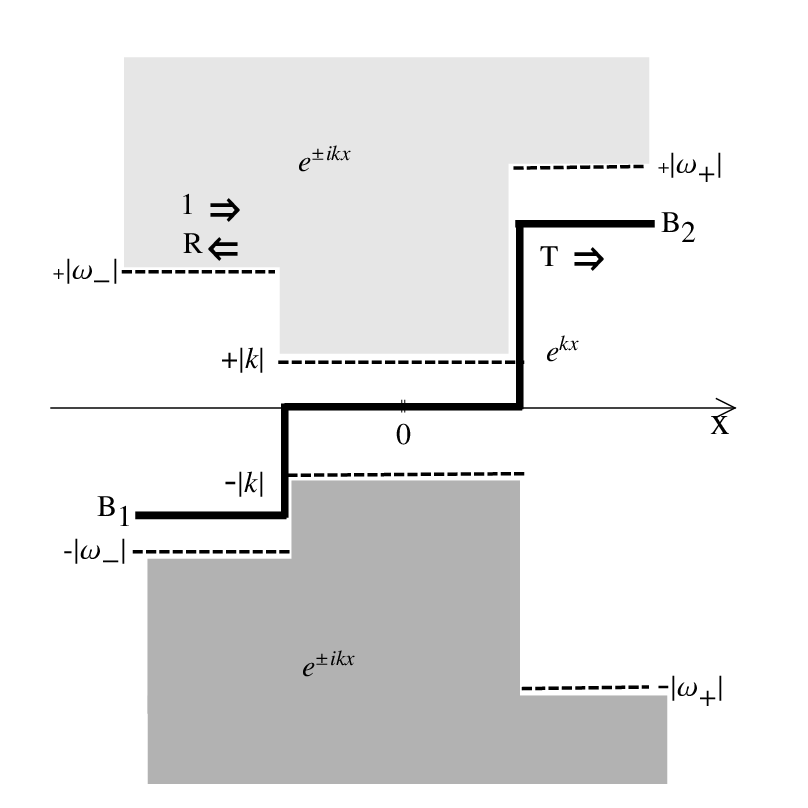}
\end{center}
{\sf Fig. 8}:  {\sf{The potential configuration associated with the double step pseudoscalar potential. Dark/light grey areas correspond to negative/positive energy continuum. The white areas correspond to bound states where the solutions are exponentials.}}\\

\begin{center}
  \includegraphics[width=3.0in]{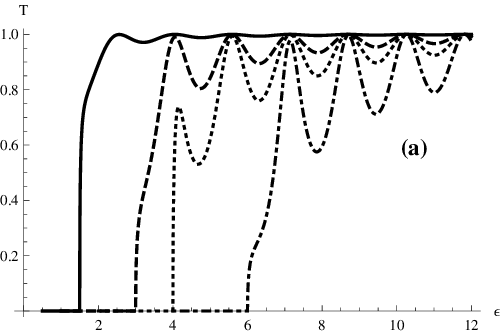}\ \ \ \ \ \ \ \
  \includegraphics[width=3.0in]{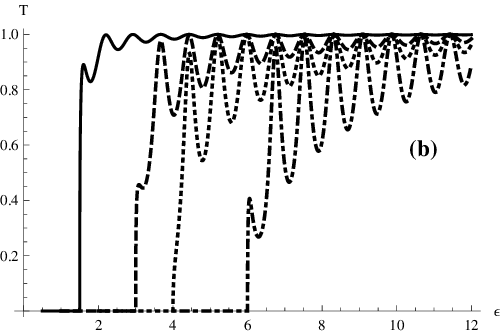}
  \end{center}
{{\sf Fig. 9}:  {\sf{The transmission coefficient as a function of energy for $B_{2}=-B_1=0.5, 2.0, 3.0, 5.0$ (left to right curves) with $k=1$. (a) for $a=1$ and (b) for $a=2$.}}}\\

\begin{center}
  \includegraphics[width=3.0in]{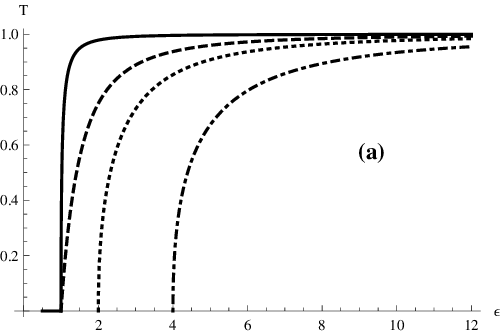}\ \ \ \ \ \ \ \
  \includegraphics[width=3.0in]{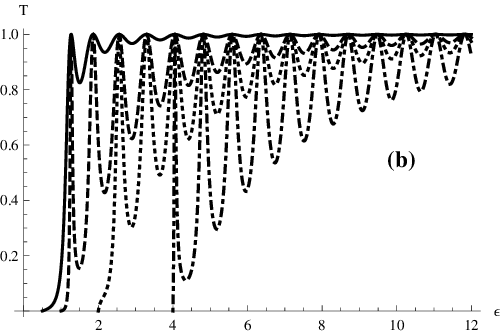}
  \end{center}
{\sf Fig. 10}:  {\sf{The transmission coefficient as a function of energy for $a=2$ and $k=1$. (a) for the Dirac particle scattered by a pseudo-scalar one step potential when $B_2=0$ and $B_{1}=-0.5, -2.0, -3.0, -5.0$ (left to right curves). (b) for our the case of $B_2=B_1=-0.5, -2.0, -3.0, -5.0$.}}\\


We wish to study the bound states of a double-step pseudoscalar potential for the case $\left|\varepsilon \right|<\left|\omega _{-} \right|$ and for $B_1=B_2$. From equation (\ref{41}) we must have $A_{-}=F_{-}=0$ in order to obtain normalizable wavefunction in the region $\left|x \right|>a/2$. We now have the full solution, given our assumption of particles incident from the left
\begin{eqnarray}\label{}
\varphi=
\left\{
  \begin{array}{ll}
    {B_-e^{{\rm }\, \kappa _{-} x} \left(_{\alpha _{-}}^{1} \right),
\qquad \qquad \qquad  \qquad \qquad  x<-a }, \\
    {{C_-}e^{{\rm i}\, k _{2} x} \left(_{\gamma _{+} }^{1} \right)+D_-e^{-{\rm i}\, k _{2} x} \left(_{\gamma _{+}^{*} }^{1} \right),
\qquad \ \ \left|x\right|<a }, \\
    {{E_-}e^{-{\rm }\, \kappa _{+} x} \left(_{\alpha _{-} }^{1} \right),
\qquad \qquad \qquad  \qquad \qquad   x>a }.
  \end{array}
\right.
\end{eqnarray}
This means that the solutions separate into even parity and odd parity states. We could have guessed this from the potential.
The even states $(C_-=D_-, B_-=E_-)$ have the (quantization) constraint on the energy that
\begin{equation}\label{cd}
\sqrt{1-\upsilon^{2}-\frac{ak}{\varepsilon}\upsilon}+\upsilon+\frac{ak}{\varepsilon}=\tan(\varepsilon)
\end{equation}
and the odd states $(C_-=-D_-, B_-=-E_-)$ have the constraint
\begin{equation}\label{dc}
\sqrt{1-\upsilon^{2}-\frac{ak}{\varepsilon}\upsilon}+\upsilon+\frac{ak}{\varepsilon}=-\cot(\varepsilon)
\end{equation}
where we have used $\upsilon=\frac{Ba}{2\varepsilon}$. These are transcendental equations, so we will solve them graphically. The plot in figure 11 shows the left hand side of the transcendental equation along with the tangent for even parity states and minus cotangent for odd parity states, respectively. The intersection of these curves gives the bound states. From  this figure we see that there are two bound states for even parity solution and two for odd parity solutions. The wider and deeper the step, the more bound states become available.
\begin{center}
  \includegraphics[width=3.5in]{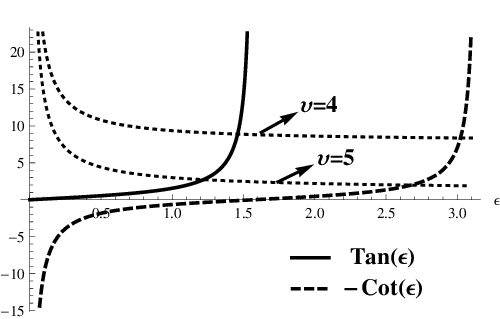}
\end{center}
{\sf Fig. 11}:  {\sf{The double-step pseudoscalar potential bound states for $k=1$ and $a=1$. The solid line represents $\tan(\varepsilon)$ and the dashed line represents $-\cot(\varepsilon)$. The dotted lines represent the left-hand side of equations (\ref{cd}) and (\ref{dc}).}}

{On the other hand, it will be interesting to underline the energy system behavior. In this end,
we plot the dispersion curves $\epsilon_k$ to show its structure in Fig 12}

\begin{center}
  \includegraphics[width=2.1in]{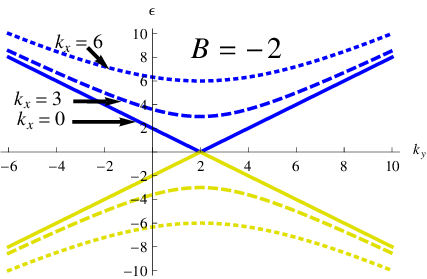}\includegraphics[width=2.1in]{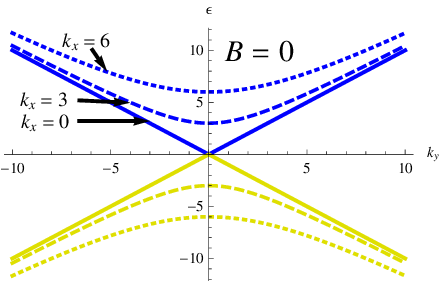}\includegraphics[width=2.1in]{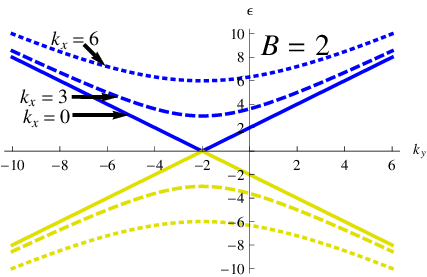}
\end{center}
{{\sf Fig. 12}:  {\sf{Dispersion curves for the oscillatory states for different values of $k_x$ and magnetic field $B$,
with $k_x$ is the wave vector along the $x$ direction ($k_x=k_-$ or $k_+$).}}

\section{Conclusion}

We considered Dirac
fermions in graphene nanoribbon subject to a sharply localized magnetic field.
After setting the theoretical model we obtained the {eigenvalues as well as the eigenspinors.} 
for a sharply localized $\delta$-type magnetic field. After matching the wavefunction at the point
$x=0$, the reflection and transmission amplitudes {were} derived in standard way.
 A study of the transmission coefficient as a function
of the strength of field $B_0$ was done, the transmission is unity
for $B_0=0$ and reduces sharply for strong field.

The general case of a bell shape magnetic field normal to the graphene sheet, namely $\vec B(x)= \hat z B_0\sech^2(\mu x)$
was studied, $B_0$ being the strength of the field and $\mu^{-1}$ stands for its range. It was shown that the solution
space of the corresponding Hamiltonian splits into three regions depending on the sign of $kB_0$ and
the energy range, $k$ is the wave vector along the $y$-axis. To obtain the reflection and transmission coefficients
we used the current density, our results reduce to those obtained for $\delta$-type magnetic field
for large values of $\mu$, here we took $\mu=1000$. Moreover, the phase shift that characterizes the transmission amplitude is given in terms of the energy and magnetic field.

Subsequently, we considered the case of an asymmetric double step pseudoscalar potential of the form
$W(x)=B_1\left[1-\Theta(x+a)\right]+B_2\Theta(x-a)$. After obtaining the general solution of the Dirac equation,
we matched the wavefunction at the points $\pm a$ to obtain the reflection and transmissions amplitudes
for the negative and positive energy solutions. Different field configuration were considered
but the interesting situation obtains when  $B_2=-B_1= 0.002$ which enables us to recover
 the reflection and transmission
coefficients for scattering of Dirac fermions by a double step pseudoscalar potential. Moreover,
for the present potential we studied  the corresponding bound states using  graphical solution
of the transcendental equations. We reached the conclusion that there are at most two bound states for even
parity and two for old parity solutions.

Finally a double Gaussian type magnetic field was considered in appendix A where a second order
differential equation for spinor components was solved {analytically}. The corresponding general solution was given in terms of
the HeunG functions. The asymptotic behavior as $x\rightarrow \pm \infty$ of these wavefunctions were {obtained and enabled} us to extract the corresponding reflection and transmission coefficients.

\section*{Acknowledgments}

The generous support provided by the Saudi Center for Theoretical Physics (SCTP)
is highly appreciated by all Authors. AJ and AE acknowledge partial support
by King Faisal University and KACST, respectively. We also acknowledge the support of KFUPM
under the theoretical physics research group project RG1108-1-2.

\section*{Appendix A: Dirac fermions in a double Gaussian type magnetic field}

We consider a magnetic field normal to the graphene sheet and of a Gaussian shape. Specifically, we take
$\vec{B}(x)=\hat{z}\left[B_1 \sech^{2}\left[\mu(x+a)\right] + B_2 \sech^{2}\left[\mu(x-a)\right]\right] $
{(Fig. 13)}, with $B_{1}$ and $B_2$ are the strengths of the field. The corresponding vector potential {(Fig. 13)} is
\begin{equation}
 \vec {A}(x)=\hat{y}\left[\frac{B_1}{\mu}\left[\tanh \left[\mu(x+a)\right]-1\right]+ \frac{B_2}{\mu}
 \left[\tanh \left[\mu(x-a)\right]+1\right]\right].
 \tag{A1} \lb{pot}
\end{equation}
We note that the double delta magnetic field considered in  section 5 is obtained in the limit $\mu \gg 1$ in the present more general model.

By using the same change of variable as in section 4 for $\pm x>0$, equation (\ref{GrindEQ6}) {can be written} in the following form
\beq
\left[4 \mu^{2} z \frac{d}{dz}\left(z\frac{d}{dz}\right) \mp
 \left({\frac {4B_{{1}}qz}{ \left( qz-1 \right) ^{2}}}+{\frac {
4B_{{2}}qz}{ \left( z-q \right) ^{2}}}\right)- \left( {\frac {2B_{{1}}}{\mu\,
 \left( qz-1 \right) }}+{\frac {2B_{{2}}z}{\mu\, \left( z-q \right) }}+
k \right) ^{2}+{\varepsilon}^{2} \right]\varphi ^{\pm } =0. \tag{A2}   \lb{qqqq}
\eeq
for $x<0$ with $q=e^{2a}$.\\

\begin{center}
  \includegraphics[width=1.6in]{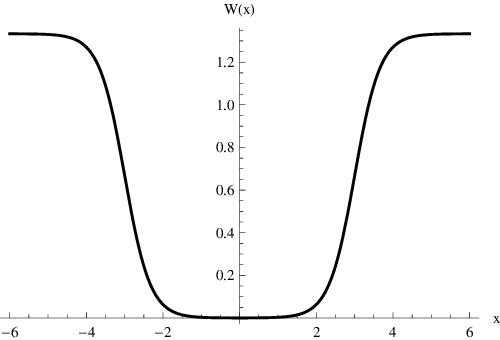}
  \includegraphics[width=1.6in]{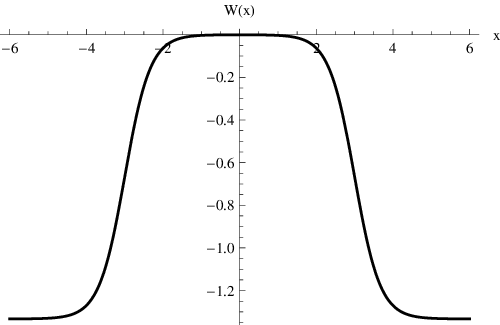}
  \includegraphics[width=1.6in]{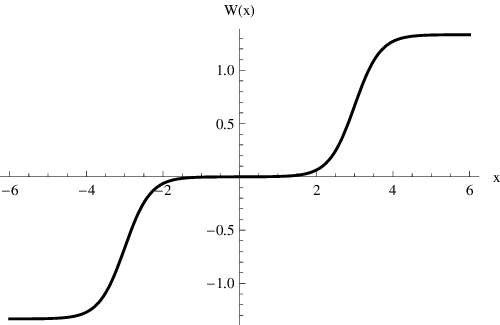}
  \includegraphics[width=1.6in]{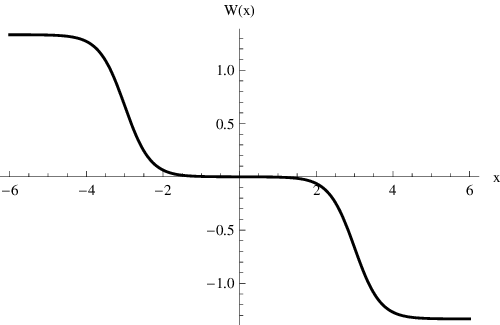}\\
  \includegraphics[width=1.6in]{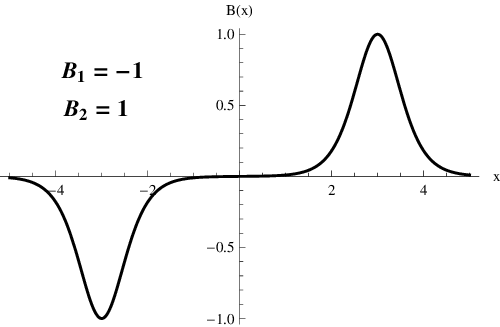}
  \includegraphics[width=1.6in]{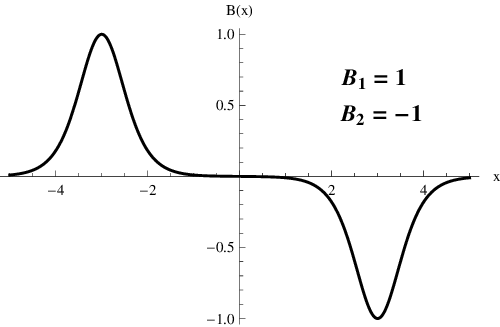}
  \includegraphics[width=1.6in]{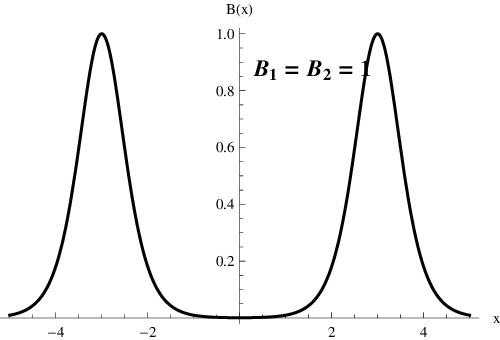}
  \includegraphics[width=1.6in]{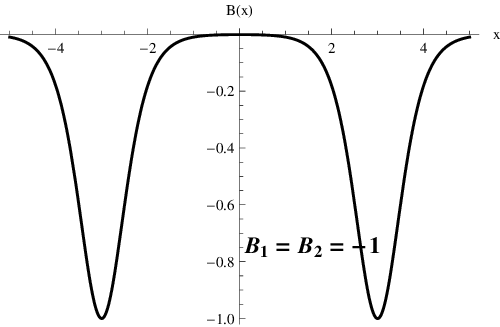}
  \end{center}
{\sf Fig. 13}:  {\sf{The pseudoscalar potential $W(x)$ and the associated double Gaussian type magnetic field $B(x)$ (\ref{pot}) for $\mu=1.5$ and $a=3$.}}\\

The general solution of (\ref{qqqq}) can be expressed in terms of the HeunG function~\cite{hf} as follows
\begin{align}
\varphi ^{+} &=A z^{l_{+}} (q-z)^{n_{+}}(-1+qz)^{m_{+}} {\mbox{HeunG}}
\left(p,b_{+},\alpha_{+},\beta_{+},\gamma_{+},\delta_{+};\frac{z}{q}\right) \tag{A3}\lb{os}\\ 
 &+B z^{-l_{+}} (q-z)^{n_{+}}(-1+qz)^{m_{+}}
{\mbox{HeunG}}\left(p,b_{1_{+}},\alpha_{1_{+}},\beta_{+},2-\gamma_{+},\delta_{+};\frac{z}{q}\right).\nonumber 
\end{align}
where we have set
\begin{eqnarray}
&& l_{\pm} =(\gamma_{\pm}-1)/2, \qquad n_{\pm}=- \delta_{\pm}/2, \qquad m_{\pm}= \left(1+\frac{S_{4\pm}}{S_{3\pm}}\right)/2,\qquad p={q}^{-2}\nonumber\\
&&  \alpha_{\pm} = \left(S_{2\pm}+S_{3\pm}\right) {\mu}^{-2}/2
,\qquad \alpha_{1\pm} = \left( -4B_{{2(1)}}+2{\mu}^{2}-S_{2\pm}+S_{3\pm}\right) {\mu}^{-2}/2,\nonumber\\
&&  \gamma_{\pm} =1 +  \mu^{-2} \sqrt {{\mu}^{2}{k}^{2}-{\mu}^{2}{\epsilon}^{2}\mp 4\,B_{{1(2)}}k\mu+{4B_{{1(2)}}}^{2}}
,\qquad \delta_{\pm}=- 2\mu^{-2}B_{2(1)} \nonumber\\
&& \beta_{\pm} =
\left[S_{2\pm} S_{3\pm}+ {\mu}^{4}+ \left( -{
\epsilon}^{2}+{k}^{2}-4\,B_{{1(2)}} \right) {\mu}^{2}\pm 4\,B_{{2(1)}}k\mu\pm{4B_{
{2}}}^{2}\mp{4B_{{1}}}^{2} \right] {\frac {\mu^{-2}}{2S_{3\pm}}} \nonumber \\
 && b_{\pm} =\mp1/2 \Bigg[  \Big[  \left( \mp{\mu}^{2}{q}^{2}\pm 2B_{{2(1)}} \right) (1-\gamma_{\pm}){\mu}^{2}
\mp{\mu}^{4}{q}^{2}\mp 2B_{{1(2)}}{q}^{2}{\mu}^{2}+k \left( 2B_{{1(2)}}{q}^{2}+2B_{
{2(1)}} \right) \mu \nonumber\\
  && \mp {2B_{{1(2)}}}^{2}{q}^{2}\mp 4B_{{1}}B_{{2}} \Big] \mu\,
S_{3\pm}\mp
 \left( S_{1\pm}+ \left( 2B_{{1(2)}}+{\mu}^{2}
  \right) ^{2}\mu \right) {\mu}^{2} \gamma_{\pm} {q}^{2}\Bigg]{\frac {{\mu}^{-5}{q}^{-2}}{S_{3\pm}}}\nonumber\\
 && b_{1\pm} = \mp 1/2 \Bigg[ \mu \Big]  \left( \pm{\mu}^{2}{q}^{2}\mp 2B_{{2(1)}}
 \right) (1-\gamma_{\pm}){\mu}^{2}\mp{\mu}^{4}{q}^{2}\mp 2B_{{1(2)}}{q}^{2}{\mu}^{2}+k \left( 2B
_{{1(2)}}{q}^{2}+2B_{{2(1)}} \right) \mu \nonumber\\
 && \mp{2B_{{1(2)}}}^{2}{q}^{2}\mp 4B_{{1}}B_{{2}} \Big] S_{3\pm}-
( 2-\gamma_{\pm} )  \left( S_{1\pm}+ \left( 2B_{{1(2)}}+{\mu}^{2} \right) ^{2}\mu \right) {\mu}^{2}{q}^{2}\Bigg]
\frac {{\mu}^{-5}{q}^{-2}}{S_{3\pm}}\nonumber
\end{eqnarray}
with
\begin{eqnarray}
S_{1\pm}&=&\sqrt { \left( 2B_{{1(2)}}+{\mu}^{2} \right) ^{2}
 \left[  \left( k+\epsilon \right) \mu\pm 2B_{{2(1)}} \right] {\mu}^{2}
 \left[  \left( k-\epsilon \right) \mu\pm 2B_{{2(1)}} \right] }\nonumber \\
 S_{2\pm} &=& -2B_{{2(1)}}+{\mu}^{2} \gamma_{+(-)} \nonumber \\
S_{3\pm} &=& \sqrt {{2\,S_{1\pm}+{\mu}^{4}+ \left( -{\epsilon}^{2}+4\,B_
{1(2)}+{k}^{2} \right) {\mu}^{2}\pm 4\,k{\mu}B_{2(1)}+4B_{2(1)}
^{2}+4B_{1}^{2}}} \nonumber \\
S_{4\pm} &=&\left( 2B_{{1}}+
{\mu}^{2} \right)  \left( {\mu}^{2}\gamma_{\pm}+2B_{{1}} \right)\nonumber.
\end{eqnarray}
\noindent The asymptotic behavior of (\ref{os}) when $x\rightarrow-\infty$  corresponds to $z\rightarrow 0$ and gives for $\varphi ^{+}$ and $\varphi ^{-}$
\begin{align}
\mathop{\lim }\limits_{x\to - \infty }\varphi ^{+} &= A e^{ik_{-} x}+B e^{-ik_{-} x}\nonumber\\
\qquad \mathop{\lim }\limits_{x\to - \infty }\varphi ^{-} &= A \rho_{+} e^{ik_{-} x}+B \rho_{-} e^{-ik_{-} x}\tag{A4}\lb{sss}
\end{align}
with
\beq
k_{-}=i\mu \sqrt{\left|\omega_{-}^{2}-\varepsilon^{2}\right|}, \qquad
\rho_{\pm}=\frac{\omega_{+}\pm ik_-}{\varepsilon}, \qquad
\omega_{+}=k+\frac{2B_2}{\mu}, \qquad \omega_{-}=k-\frac{2B_1}{\mu}\nonumber.
\eeq

 In the region when $x>0$ the analogous equation to (\ref{GrindEQ6}) reads
\beq
\left[4 \mu^{2} z \frac{d}{dz}\left(z\frac{d}{dz}\right) \mp
 \left({\frac {4B_{{1}}qz}{ \left( q-z \right) ^{2}}}+{\frac {
4B_{{2}}qz}{ \left( 1-qz \right) ^{2}}}\right)- \left( {\frac {2B_{{1}}z}{\mu\,
 \left( q-z \right) }}+{\frac {2B_{{2}}}{\mu\, \left( 1-qz \right) }}+
k \right) ^{2}+{\varepsilon}^{2} \right]\varphi ^{\pm } =0.\tag{A5}   \lb{qg}
\eeq
The solution of (\ref{qg}) can be expressed in terms of the HeunG function as follows
\begin{align}
\varphi ^{+} &=C z^{l_{-}} (q-z)^{n_{-}}(-1+qz)^{m_{-}} {\mbox{HeunG}}\left(p,b_{-},\alpha_{-},\beta_{-},\gamma_{-},\delta_{-};\frac{z}{q}\right)\tag{A6}\lb{oDs}\\
 &+D z^{-l_{-}} (q-z)^{n_{-}}(-1+qz)^{m_{-}}
{\mbox{HeunG}}\left(p,b_{1_{-}},\alpha_{1_{-}},\beta_{-},2-\gamma_{-},\delta_{-};\frac{z}{q}\right)\nonumber
\end{align}
where the wavefunction parameters have been defined previously. The
asymptotic behavior of (\ref{oDs}) when $x\rightarrow +\infty$, which corresponds to $z\rightarrow 0$, gives for $\varphi ^{+}$ and $\varphi ^{-}$
{while the} asymptotic behavior of (\ref{os}) when $x\rightarrow-\infty$  corresponds to $z\rightarrow 0$ and gives for $\varphi ^{+}$ and $\varphi ^{-}$
\begin{align}
\mathop{\lim }\limits_{x\to + \infty }\varphi ^{+} &=C e^{-ik_{+} x}+D e^{ik_{+} x}\nonumber\\
\qquad \mathop{\lim }\limits_{x\to + \infty }\varphi ^{-} &=C \rho e^{-ik_{+} x}+D \rho e^{ik_{+} x}\tag{A7}\lb{ssss}
\end{align}
with $k_{+}=i\mu \sqrt{\left|\omega_{+}^{2}-\varepsilon^{2}\right|}$ and $\rho=\frac{\omega_{+}+ ik_+}{\varepsilon}$. The reflection and transmission coefficients for the positive and negative energy solutions are given by equation (\ref{wx}){, with} the reflection and the transmission amplitudes given by $r=\frac{\rho_{2}-\rho}{\rho-\rho_1}$ and $t=\frac{\rho_2-\rho_1}{\rho-\rho_1}$. {Figure 14} show transmission for the Dirac particle scattered by the pseudo-scalar potential (\ref{pot}), we can see from figures 13 (a) and (d) that, the EWs, TWs and BS 
appear in the energy ranges $\left|\omega_-\right|<\left|\varepsilon\right|<\left|\omega_+\right|$, $\left|\varepsilon\right|>\left|\omega_+\right|$ and $\left|\varepsilon\right|<\left|\omega_-\right|$, respectively. In this later case the pseudo-scalar potential (\ref{pot}) behaves like a potential {well}. {In figures 14} (b) and (c) the pseudo-scalar potential (\ref{pot}) behaves like a potential barrier. We also notice the vanishing of the energy range for the EWs  but the TWs and BS appear for the energy $\left|\varepsilon\right|>\left|\omega_-\right|$ and $\left|\varepsilon\right|<\left|\omega_-\right|$, respectively. {The energy ranges for EWs, TWs and BS are} affected by the value of the parameter $\mu$.\\

\begin{center}
  \includegraphics[width=1.6in]{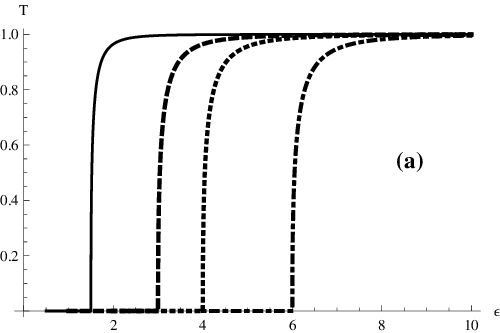}
  \includegraphics[width=1.6in]{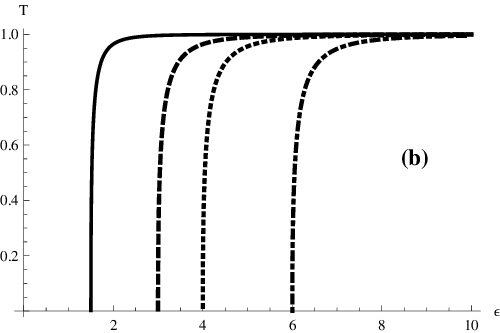}
  \includegraphics[width=1.6in]{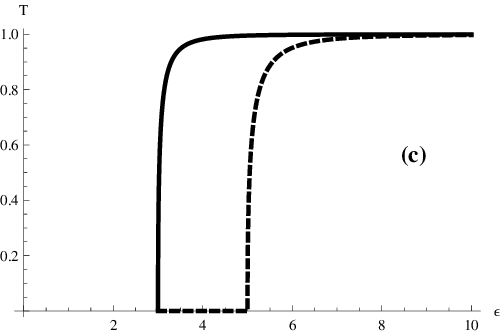}
  \includegraphics[width=1.6in]{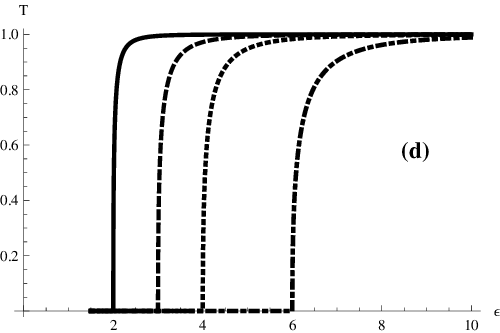}
\end{center}
{\sf Fig. 14}:  {\sf{The transmission coefficient as a function of energy. (a) $2B_1=2B_{2}=0.5, 2.0, 3.0, 5.0$ (left to right curves). (b) $2B_1=2B_2=-0.5, -2.0, -3.0, -5.0$. (c) $2B_1=-2.0$ and $2B_2=-3.0, -6.0$. (d) $2B_1=-0.5$ and $2B_2=1.0, 2.0, 3.0, 5.0$. With $a=1$, $k=1$ and $\mu=1$. Note that, for the case $B_1=-B_2$, we have total transmission.}}


\begin{thebibliography}{1}
\bibitem{15} The non-physical fields (i.e., gauge fields) could be eliminated by gauge transformation.
\end{thebibliography}
\end{document}